# Optical Properties of Chiral Graphene Nanoribbons: a First Principal Study

M. Berahman[1, 2†‡], M. Asad[1], M. H. Sheikhi[1, 2]

[1] Nanotechnology Research Institute, Shiraz University, Shiraz, Iran

[2] School of Electrical and Compotator Engineering, Shiraz University, Shiraz, Iran

[†] berahman@shirazu.ac.ir

[‡] Corresponding author

**Abstract**:

In this paper, optical properties of Chiral Graphene Nanoribbons both in longitude and transverse polarization have been studied using density functional theory calculation. It has been shown that the selection rule which have been reported before for Armchair and Zigzag Graphene Nanoribbons are no longer valid due to breaking symmetry on these new categorize of graphene nanoribbons. However, still the edge states play a critical role in optical absorption. It have been illustrated that depending on the polarization of incident beam the absorption peaks are different while it is spread in the same energy range. It is also suggested that the absorption of light is sensitive to the chiral vector on the edges and direction of the light polarization. Due to breaking symmetry in chiral graphene nanoribbons, absorption peak is changed and it would be around 1000nm, introducing a new potential of graphene nanoribbons for optoelectronic devices.

## 1. Introduction

Graphene, the 2 dimensional structure of graphite, have shown lots of interesting Electrical [1], magnetic [2] and transport [3] properties since its synthesis in 2004 [4]. Very high electron mobility of this material, which is due to Dirac particles, makes the Graphene almost unaffected from different kind of scattering such as electron-phonon or electron- electron scatterings [1, 3]. These properties make Graphene good candidate for electrical engineering usage [5-9]. Beside these advantages, having no band gaps make Graphene noise-effective. Confining one of the dimensions of Graphene would solve this problem. These 1 dimensional allotropes of carbon are known as Graphene Nanoribbons (GNRs).

Recently, GNRs with few defects crystallography and perfect edge structures have successfully synthesized using AFM cut [10] or longitude unzipping of carbon Nanotubes (CNTs) [11-14]. GNRs have technologically promising electronic [15, 16], transport [16-20] and optical properties [21-28] due to confinement of electrons with finite width. They show very similar transport properties while have small band gaps that make them tunable by using gated configurations.

One of the most fascinating GNRs' features caused by of quantum confinement is sensitivity to edge geometrical topology [15-21]. Generally GNRs have been categorized into zigzag and Armchair edges [15-21]. However, recent success in unzipping longitude CNTs [29] showed that it would be possible to synthesis another kind of GNRs with different edge confinement and electrical characteristics [29-31]. These new family of GNRs, have edge structure between Zigzag and Armchair, though, named as Chiral GNRs. While chiral GNRs (CGNRs) have low geometrical symmetry, Armchair and Zigzag edged are known as high symmetrical. AGNRs and ZGNRs can be classified by the number of dimmer lines or zigzag chins across the ribbons width. Chiral, on the other hand, have been classified using both chiral vector and widths.

Optical properties of Armchair and Zigzag GNRs have been studied before [21-28]. In ZGNRs, most local peaks in optical spectrum have mixed polarization characteristic. This is because of the fact that for magnetic ground states, the reflection about yz plane is broken, leading to mixed polarization [27]. It have been shown before that optical transition in Z direction of ZGNRs always happen between same parity of states [22,24] and are dominant when the difference between the sub bands numbers are $\pm 1$ [27] while y direction have direct transition [27]. On the other hand, these forbidden transitions for ZGNRs are reported elsewhere [25]. It has been reported that, owning to the symmetry of AGNRs, peaks corresponding to Z and Y polarized photons are well separated in energy [28]. Transition in Z direction is happen between odd parity states and direct band gaps [23] while it is different in other reports [27]. Different families of AGNRs can show different optical characteristic from simple dominate low energy peak in semi-metals to more transition peaks in semiconductors [25]. The absorption frequency spread from low energies (below 0.5ev) to high energies (approximately 5ev) while dominant peaks are tolerated between 0.5 to 2.5 ev. However all previous reports, particularly in Zigzag and Armchair edges, show the importance of edge states on optical properties of GNRs [21-28]. These theoretical studies demonstrate the capability of controlling optical absorption in GNRs with width and edges and already show the potential of this material for future nano electro optic engineering devices. However, introducing chiral vector could make GNRS into new level of capability for better devices with more accurate selectivity in optical absorption spectrum.

CGNRs have interesting magnetic and electrical properties. Electrical and magnetic properties of CGNRs have been studied before using tight binding and ab initio [29-31]. It has been shown that these properties critically depend on the percentage of carbon atoms at the zigzag sites [31]. It has been reported both theoretically and experimentally that increasing width could decrease band gaps of CGNRs [29]. Also it has been demonstrated that magnetic moments and edge state energy splitting are dependent on width and chiral angle [30]. However, as far as we know, there is no report on optical properties of chiral GNRs. In this paper, optical properties of different family of CGNRs have been studied.

The rest of this paper is organized as follows. In section 2 geometrical characteristic of CGNRs have been introduced and computational method is scrutinized. Section 3 contains result and discussion of dielectric and absorption spectrum. Finally, in section 4 main results would be summarized.

## 2. Computational Method

Unzipping CNTs along its length direction could induce Chiral GNRs. These CGNRs could be categorized using chiral vectors (n, m) and width (w), where n>m and both n and m are natural numbers and w is known as the width and is defined as the number of dimmer atoms along the Armchair direction as shown in figure 1. It could be possible to use the chiral angle instead of chiral vector which defined as degree between chiral vector and zigzag direction and is calculated as follows:

$$\theta = \arcsin\sqrt{\frac{3m^2}{4(n^2+mn+m^2)}} \qquad (1)$$

In our model all the CGNRs are terminated by the Hydrogen atoms in order to omit dangling bands effects. In order to demonstrate important angles of CGNRs, (n, 1) angles use as models, where n<7. The length of the carbon- carbon bond is chosen as 1.24A and the length of carbon-hydrogen bond is 1.101A.

The ground state electronic properties of the relaxed system obtained by performing DFT calculation, within the LDA with PZ functional for the exchange and correlation effects of the electrons as implemented in the Quantum Espresso suite of programs [32]. Optimization of each structural configuration is performed by relaxation of all atomic positions to minimize their total energies until the maximum force on any atom was less that 0.05eV/A. The electron-ion interaction was described by the ultra soft pseudo potential, and the energy cut off was set to be 2000ev. Self consistent field procedures were performed with a convergence criterion of $10^{-5}$ a.u. on the energy and electron density.

To calculate the macroscopic dielectric, self consistent field used in which each electron interacts independently with a self consistent electromagnetic field [33]. Dielectric constant has been calculated based on wave function results of Hamiltonian diagonalization matrix. It defined as follow [33]:

$$\frac{1}{\varepsilon(\omega)} = \left[(1+T)^{-1}\right] \qquad (2)$$

Where $\varepsilon(\omega)$ is dielectric constant within frequency region. Elements of the matrix T is $T_{k,k'}$ and defined by:

$$T_{kk'} = \frac{4\pi e^2}{(q+k)^2} \sum_{kll'} \frac{f_0(E_{k+q},l) - f_0(E_{kl})}{E_{k+q,l'} - E_{kl} - \hbar\omega + i\Gamma} \eta_k^* \eta_{k'} \qquad (3)$$

Where q is restricted to lie in the first Brillouin Zone and the K's are reciprocal lattice vectors. $\eta_k$ is define as wave function for K vector which calculated using previous results from DFT. $f_0(E)$ is Fermi-Dirac destruction function. E is electron charge. $\Gamma$ is small energy which would be defined. $\hbar\omega$ is energy incident beam. For long wavelength limit, it have been shown that equation (2) would be reduced to [33]

$$\varepsilon(\omega) = 1 + T_{00} + \sum_{\substack{k \\ k \neq 0}} T_{0K} \left[ (1+T)^{ok} \Big/ (1+T)^{00} \right] \qquad (4)$$

Where the superscript denotes the cofactor of the matrix elements. Equation 4 has been used in order to calculate both imaginary and real part of dielectric constant. These calculations have been done with broadening equal to 0.1ev for 8 bands in conduction band and 8 bands in valance band. Optical absorption spectrum can be calculated from dielectric constant by its famous equation [34]:

$$\alpha(\omega) \approx \frac{\omega}{\sqrt{\varepsilon_2(\omega)}} \qquad (5)$$

Where $\varepsilon_2$ is imaginary part of dielectric constant.

### 3. Results and Discussions

Figure 2 shows Bloch wave function for the (5, 1) chiral GNR in $\Gamma$ and Z points in Brillouin zone. As can be seen, there is a distinct difference between valance and conduction band Bloch functions. The occupied valance band states appear to be connected along the nanoribbon direction, while the unoccupied conduction band states are more oriented across the nanoribbon. At $\Gamma$ point, the states are localized towards the edges of the chiral nanoribbon. This has been shown the importances of edge states in nanoribbons are valid for CGNRs, too.

Figure 3 shows imaginary part of dielectric constant versus energy for some samples of CGNRs. $M_{ij}$ denotes a peak in the spectrum due to a transition from the ith valance band (counted from the top) to the jth conduction band (counted from the bottom). As show in figure 3, the (1, 0) CGNR which is equal to the Zigzag GNR, selection rule can be confirmed. As can be seen, $M_{11}$ is important transition in all CGNRs with n=1, and as angle decreases this become almost the dominate peaks in spectrum. These suggest that despite of ZGNRs, direct transition is possible and important in longitude polarization.

Like ZGNRs spectrum, $M_{12}$ and $M_{21}$ are important in chiral, and despite of (5, 1), they create a local maximum peak around 1.2 ev. While number of Zigzag chains is increased in CGNRs, more transitions are gathered around 1.2ev so almost dominate peak of spectrum in (5,1) is created around 1.2 ev while effect if $M_{12}$ and $M_{21}$ is vanished.

In presence of transverse polarized light, most local and nonlocal peaks are created due to direct connection, specifically from band gaps, and in this point, chiral GNRs are most likely to Zigzag and Armchairs.

Increasing number of Zigzag chains could induce selection rules of ZGNRs in transverse direction correct for CGNRs, too. In (5, 1), or more which are not plotted here, most dominate peaks are happened in odd direct transitions. ZGNRs have no absorption under 1 ev, while CNGRs have the local peak which is created by the band gap in their energy spectrum. It could be suggested that this peak is created by local localization of wave function near armchair edges. This conclusion can be scrutinized by increasing the number of Armchair chain in CGNRS, i.e. (3, 2), which decrease the localization and completely vanish this peak in armchair edges.

By following a column of atoms in usual Zigzag GNR, which is the mother of chiral unit cells, it could be clearly seen a full Graphene (two atoms A and B) in each GNR. However in chiral GNRs there is always a missed atom compared to the AGNRs unit cell which is a perspective of breaking symmetry inside Graphene unit cell and consequently in AGNRs unit cell. These asymmetry cause activation in the bond of edge structure which caused breaking in selection rules of optic in regular GNRs in some cases but in some others, the selection rules are still correct. This is exactly the reason of disappearing separation of longitude and transverse spectrum which could be seen clearly in Armchair and Zigzag GNRs. By diverging CGNRs, either by increasing the parameter m or n of the chiral vector in proper way, The transverse symmetry along the Z direction increased, so the selection rules in ZGNRs also has rule to create the most important local peaks in CGNRs.

Figure 4, shows optical absorption spectrum of chiral GNRs versus the wavelength. The red region in figure 4 shows the visible wavelength spectrum. As illustrated, CGNRs have completely different spectrum versus Armchair and Zigzag GNRs. While Zigzag have dominate peak in low wavelength and Armchair have a very high wavelength absorption, chiral have dominant absorption peaks between these two groups. As can be seen, by increasing the angle of chiral GNRs, the dominant peaks in wavelength is slightly increased in the range from below 1000nm to above it. However, almost all CGNRs with n=1, show similar spectrum in visible region. This property could be very useful in drop growth technique in optic sensors which there are no specific control on formation types of CNTs and GNRs.

## 4. Conclusions

In this paper, optical properties of the chiral edge Graphene Nanoribbons have been studied within DFT approximation and it have been compared to Zigzag and Armchair edge Graphene Nanoribbons. It has been shown that because of breaking yz plane symmetry in CGNRs, selection rules which were applied to zigzag Graphene Nanoribbons can not be applied and every transition depending on how this symmetry breaks is possible. It is shown that depending on the polarization of incident beam the absorption peaks are different while it is spread in the same energy bands.

Also, it have been illustrated that, like Zigzag and Armchair, in Chiral Graphene Nanoribbons, the edge states play a critical role in optical absorption. They are involved in almost all the important absorption peaks in the optical region. However, despite of Armchair and Zigzag, the peaks in Chiral GNRs are in 1000nm which demonstrates its power in optical Engineering devices in this range of wavelength.

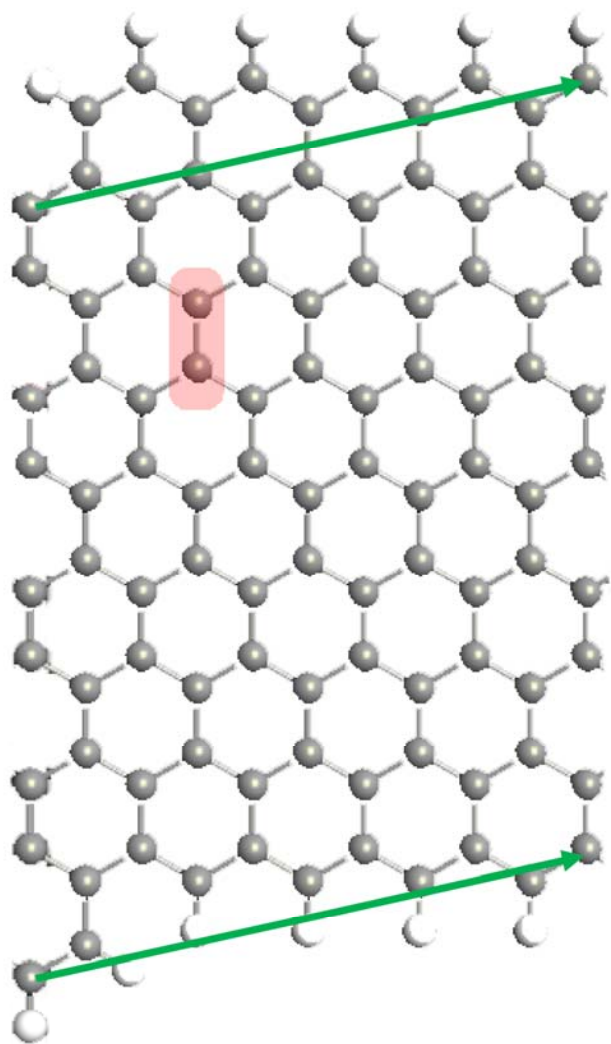

Figure 1: (5,1) Chiral Graphene Nanoribbons. Gray circles are Carbon atoms and white circles are Hydrogen atoms. The Green vector is Chiral Vector. The red area illustrates atoms in the unit cell of Graphene.

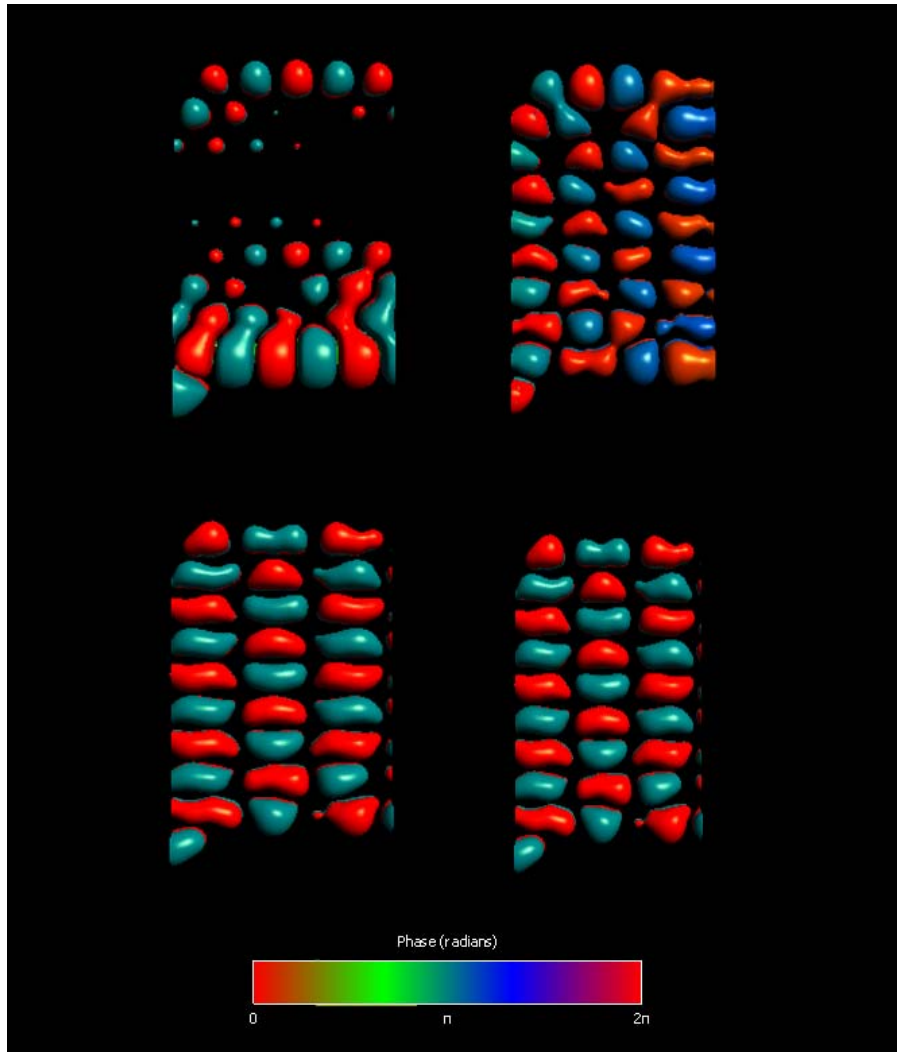

Figure 2: From left to right the Γ point and the Z point Bloch wave function of (5, 1) Chiral Graphene Nanoribbon. The top and bottom rows display the conduction and valance band, respectively.

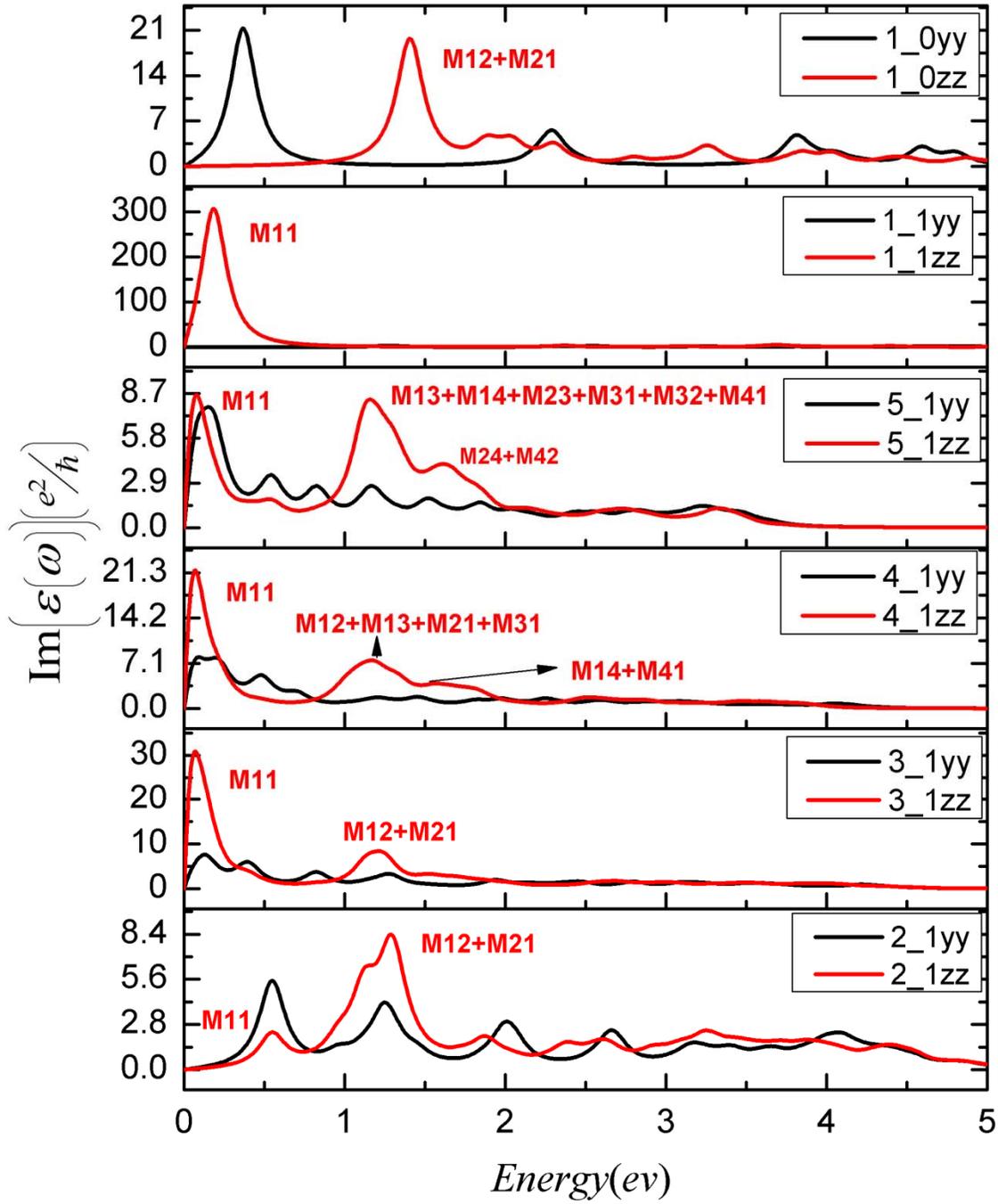

Figure 3: Imaginary part of dielectric constant versus Energy for different kind of Chiral Graphene Nanoribbons. Red curves demonstrate Z polarization of incident beam and Black curves illustrate Y polarization of incident beam

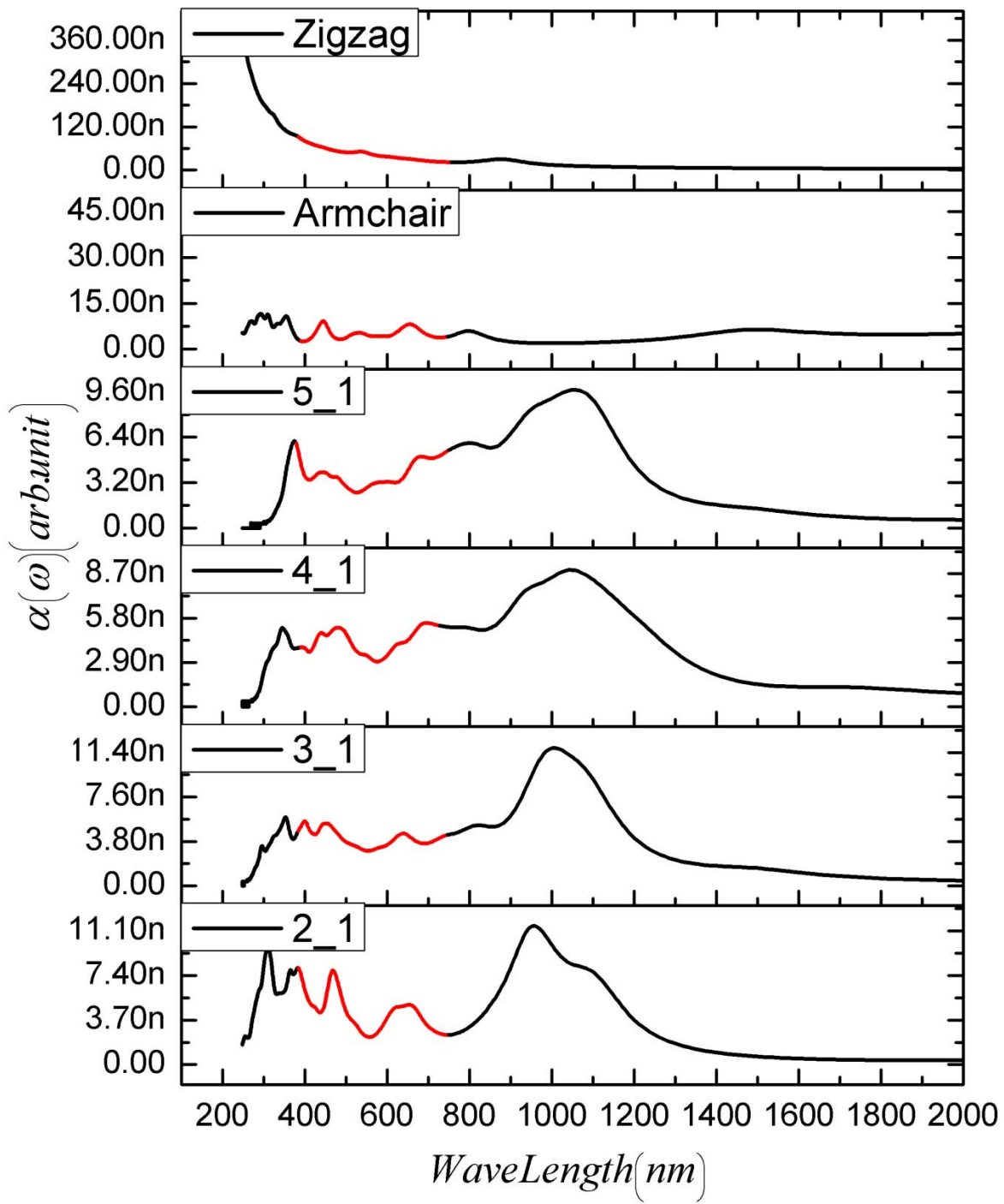

Figure 4: Optical Absorption versus Wavelength for different Chiral Graphene Nanoribbons. The Red Area illustrates visible spectrum range.